# On Basics of Cosmology

S N Pandey and B K Sinha
1-MNNIT, Allahabad(U.P.)-India
2-HCST,Farah,Mathura(U.P.)-India

**Abstract:**

Some discussion of physical and geometrical interpretation of Einstein's theory of gravitation which is on basic of cosmology.

## Introduction:

The fundamental idea of geometrical theory of gravity start from the fact to assign four co-ordinates, say, (x,y,z,t) to any event observed in our vicinity. Locally space appears flat but this does not prefridge the global shape of the space.

The surface of a four-dimensional hyper-sphere:

$$(x')^2 + (x^2)^2 + (x^3)^2 + (x^4)^2 = a^2 \tag{1}$$

Where a is the radius of the sphere. The distance between any two nearby point on this surface is

$$dl^2 = (dx')^2 + (dx^2)^2 + (dx^3)^2 + (dx^4)^2 \tag{2}$$

Here it should be noted that $x^4$ has nothing to do with time. It is extra and unphysical as curved three dimensional space is a subspace of a flat Euclidian space. Eq.(1) can be used to eliminate unphysical co-ordinate $x^4$ in (2):

$$dl^2 = (dx')^2 + (dx^2)^2 + (dx^3)^2 + \frac{(x'dx' + x^2 dx^2 + x^3 dx^3)^2}{a^2 - (x')^2 - (x^2)^2 - (x^3)^2} \tag{3}$$

In this expression there appears only the physical co-ordinates x', $x^2$, $x^3$.

## The case of positive curvature:

Introduce the polar co-ordinates:

$$x' = r\sin\theta\cos\phi \qquad , x^2 = r\sin\theta\sin\phi \qquad x^3 = r\cos\theta$$

There the space interval equation (3) takes the form

$$dl^2 = \frac{dr^2}{1-\frac{r^2}{a^2}} + r^2 d\theta^2 + r^2 \sin^2\theta\, d\phi^2 \tag{4}$$

It is easy to see that rectangular coordinates $x^k$ and the radial coordinate r are actually periodic coordinates.

To have understanding of our space of positive curvature, let us work at the radius and circumference of a wide placed in this space. For convenience, consider a circle defined by r=b= const. around the origin. This circle has as radius the distance between r=0 and r=b

$$l = \text{radius} = \int_0^b \frac{dr}{\sqrt{1-r^2/a^2}} = a\sin^{-1}\left(\frac{b}{a}\right) \tag{5}$$

If the circle is in the plane $\theta = \frac{\pi}{2}$

$$\text{Circumference} = \int_0^{2\pi} b\sin\theta\, d\phi = 2\pi b \tag{6}$$

The ratio of radius to circumference is therefore

Larger than $1/2\pi$ (familiar property of spaces of positive curvature).

The surface area of the sphere r=b=const. surrounding origin is:

$$\text{Area} = \int_0^{2\pi}\int_0^{\pi} b^2 \sin\theta\, d\theta d\phi = 4\pi b^2$$

Hence the ratio of the radius squared to the area is larger than $1/4\pi$. For a radius larger than $\pi a/2$ the area decreases as radius increases. The volume inside the sphere r=b is

$$\text{Volume} = \int_0^{2\pi}\int_0^{\pi}\int_0^{b} \frac{r^2}{\sqrt{1-\frac{r^2}{a^2}}} \, dr \sin\theta \, d\theta \, d\phi$$

$$= 4\pi \left( \frac{a^2}{2} \sin^{-1}\frac{b}{a} - \frac{ba^2}{2}\sqrt{1 - b^2/a^2} \right) \tag{7}$$

To obtain the total volume of the three –sphere we must have to take

$$b=0 \, , \, \sin^{-1}\frac{b}{a} = \pi$$

The bottom of sphere. Then , volume is given by $2\pi^2 a^3$ . Our 3- sphere is a "closed space"; it has a finite volume though it has no boundaries .

Since r is a periodic coordinate , it is convenient to interoduce new angular coordinate x such that

$$r = a \sin x \qquad , \, 0 < x < \pi$$

x is a single valued , so advantageous , and we get (4) :

$$dl^2 = a^2 \left[ dx^2 + \sin^2 x + (d\theta^2 + \sin^2\theta \, d\phi^2) \right] \tag{8}$$

and    l = radius distance = ax    (9)

from (5).

Now we add the time as the fourth coordinate and we have the space time interval for closed isotropic universes as:

$$ds^2 = c^2 dt^2 - a^2(t)\left[ dx^2 + \sin^2 x (d\theta^2 + \sin^2\theta \, d\phi^2) \right]$$

If we replace time t by cosmic time $\eta$ defined by

$$cdt = a(\eta)d\eta$$

Then

$$ds^2 = a^2(\eta)\left[ d\eta^2 - dx^2 - \sin^2 x (d\theta^2 + \sin^2\theta \, d\phi^2) \right] \tag{10}$$

The Ricci tensor is entirely determined as

$$R_{00} = \frac{3}{a^2}(a\ddot{a} - \dot{a}^2)$$

$$R_{kn} = \frac{1}{a^4}(2a^2 + \dot{a}^2 + a\ddot{a})g_{kn} \qquad \left(\frac{da}{dt} = \frac{1}{a}\frac{da}{d\eta} = \frac{\dot{a}}{a}\right)$$

$$R = R_0^0 + R_k^k = \frac{6}{a^3}(a + \ddot{a})$$

The Einstein equation

$$R_i^k - \frac{1}{2}\delta_i^k R = -8\pi\, G T_i^k$$

00-component of the equation reduces to

$$-\frac{3}{a^4}(a^2 + \dot{a}^2) = -8\pi G T_0^0 \tag{11}$$

Note that in this equation $\ddot{a}$ is cancelled. Thsi equation determines a(t) if $T_{00}$ is given. The other components of Einstein equations can be regarded as ignored because they tell us nothing different from eq(11).

The main contribution towards energy density would be due to "particles" of one kind another (galaxies, hydrogen gas, etc.). We will neglect the pressure that these particles might exist. Hence

$$T_i^k = \rho u_i u^k$$

Where $\rho$ is the proper mass density. In our co-moving coordinates, the matter is at rest and hence

$$T_0^0 = \rho$$

Again we have shown

$$\rho(t) = \frac{M}{2\pi^2 a^3}$$

Where M is a constant as it is the "total mass of the universe". It is meaningless to talk of the mass of the universe from an operational point of view, M is really the sum of all the proper masses of the particles in the universe. But, such a sum will fail to account kinetic and binding energies to give the total mass. Hence we get

$$\frac{3}{a^4}\left(\dot{a}^2 + a^2\right) = \frac{4GM}{\pi a^3} \tag{12}$$

This is the differential equation that describes the closed Freidmaun model.

Its solution is,

$$a(\eta) = a^* (1 - \cos \eta)$$

$$a^* = \text{———} \tag{13}$$

$$t = a^* (\eta - \sin \eta)$$

So, this can be regarded as parametric equations for a (t), and the curve represented by these is a Cycloid:

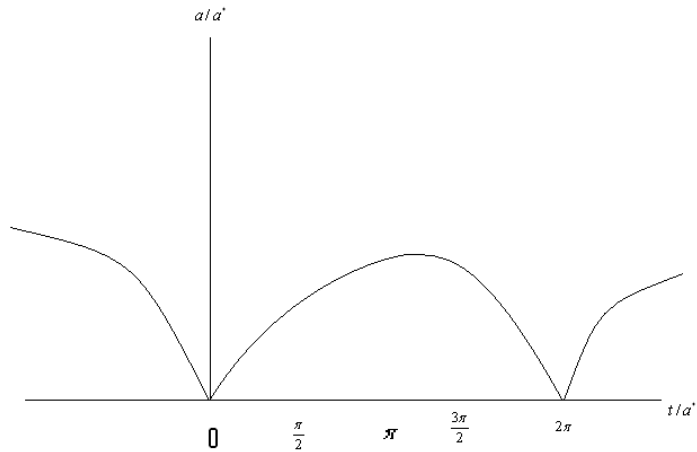

at t = 0, ± 2πa*, ±4πa*, ……………,a(t) vanishes, i.e., the universe contracts to a point. Since the density becomes very large when this is about to happen, Our approximate expression for energy- momentum tensor will fail.

Similar calculation for **Negative Curvature** for the open universe will lead to

$-(\dot{} - a^2) = 4GH/ a^3$

Or, similar $a = a^* (\cosh \eta - 1)$

$t = a^* (\sinh \eta - 1)$

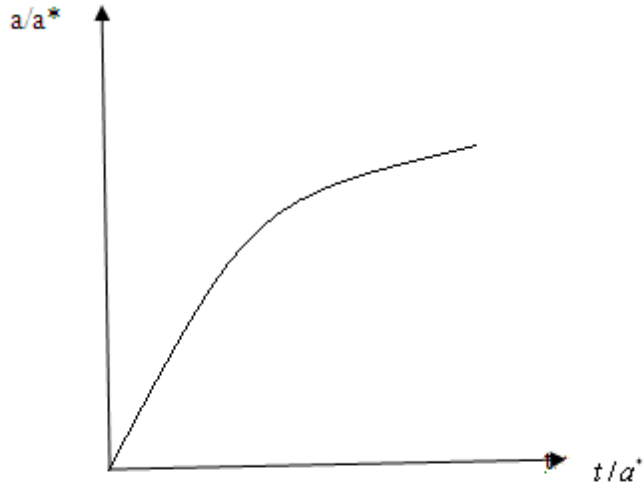

$$dt^2 = (dx_1)^2 + (dx_2)^2 + (dx_3)^2 + \frac{(\quad)}{(\ )(\ )(\ )} \tag{14}$$

$$= \frac{}{/} + \quad + \quad \sin^2 \emptyset$$

$l = $ radius $= a \sinh \quad /$

Hence, ratio of radius to circumference smaller than $1/2$.

Volume $= 4 \quad \overline{\quad} \sinh \quad - + \overline{\quad} \quad 1 + \overline{\quad}$

as $b \to \infty$, $v \to \infty$, Hence space of negative Curvature is open and INFINITE .i.e. Universe begins with bang and continuous to expand for—,as $t \to \infty$, the universe gradually becomes flat.

For **Zero Curvature**, the universe is with flat three geometry. The line element is

$$ds^2 = a^2(\eta)(\ \eta \quad - \quad - \quad - \quad \emptyset) \tag{15}$$

This space is open and infinite.

And the equitation of motion is $\dot{-} = \dfrac{-}{}$

Yielding: $\quad a(t) \propto (\dfrac{}{})^{1/} \cdot t^{2/3}$

As t→ ∞, hen four geometry leads to become flat.

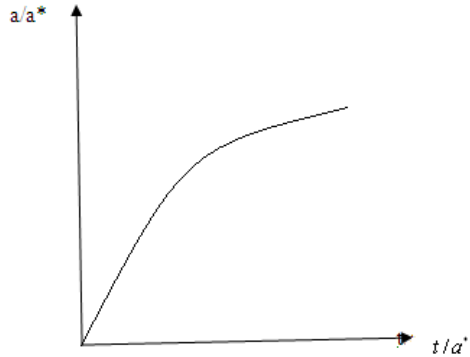

$= \quad + \quad + \quad \emptyset$

$l = r$

Volume = infinite and open

(Matter generates curvature in four geometry; it need not generate curvature in three geometry)

Now suppose we have cosmological constant $\Lambda \neq 0$, we will assume that $\Lambda \gg 8$ , then we neglect altogether and regard the universe as empty the Behaviors of the universe are three controlled by $\Lambda$. This has roughly same effect as a uniform mass density, constant in space and time. $\Lambda > 0$ corresponds to a negative effect mass density, i.e. repulsion and $\Lambda < 0$ corresponds to a positive mass density, i.e. attraction. Hence the universes with the positive values the Expression will take place and will tend to accelerate and in the universes with negative values will slow down expansion, stop and reverse. Thus, **Positive Curvature** Neglecting we obtain is $-(\dot{} + a^2) = -$ which becomes on putting $\dot{} = \dfrac{-}{}$

$(\dfrac{}{}) = -1 + (\Lambda a^2)/3 \qquad (16)$

From here, it is obvious that $-1+ (\Lambda a^2)/3$ cannot be negative. This implies that $\frac{\overline{\phantom{x}}}{\Lambda} > 0$ and the value of a can never be less than $\frac{\overline{\phantom{x}}}{\Lambda}$. In this case (Lemaitre model) with positive curvature requires a positive cosmological constant, and its radius of curvature is never zero, that is, **NO BIG BANG**. Integrating we have:

$$a(t) = \frac{\overline{\phantom{x}}}{\Lambda} \cosh \frac{\Lambda(\phantom{xxx})}{\overline{\phantom{xxx}}} \tag{17}$$

Where $t$ is the time at which a has its minimum value. For $t > t$, the universe expands monotonically, and as $t \to \infty$ the universe becomes flat.

Negative curvature here

$$\overline{\phantom{xxx}} = 1+ (\Lambda a^2)/3$$

Yielding 
$$a(t) = \frac{\overline{\phantom{x}}}{\Lambda} \sinh \frac{\overline{\Lambda}}{\overline{\phantom{x}}} \quad \text{for } \Lambda > 0 \tag{18}$$

$$a(t) = \frac{\overline{\phantom{x}}}{|\Lambda|} \sin \frac{\overline{|\Lambda|}}{\overline{\phantom{x}}} \quad \text{for } \Lambda < 0 \tag{19}$$

Both the universe begin with a bang at $t = 0$. The first one expands monotonically while second one oscillates with period

$$2 \frac{\overline{\phantom{x}}}{|\Lambda|}$$

In our actual universe mass density near big bang must have been very large and have empty Lemaitre model cannot be used to describe the behavior near $t = 0$.

**Zero Curvature**

The equation of motion is

$$\text{—} = (\Lambda\, a^2)/3 \tag{20}$$

This equation makes sense if $\Lambda > 0$; and hence its solution is usually called the de Sitter model. e.g. (21) has an exponentially decreasing solution but this is of no use or relevance to our expanding universe.

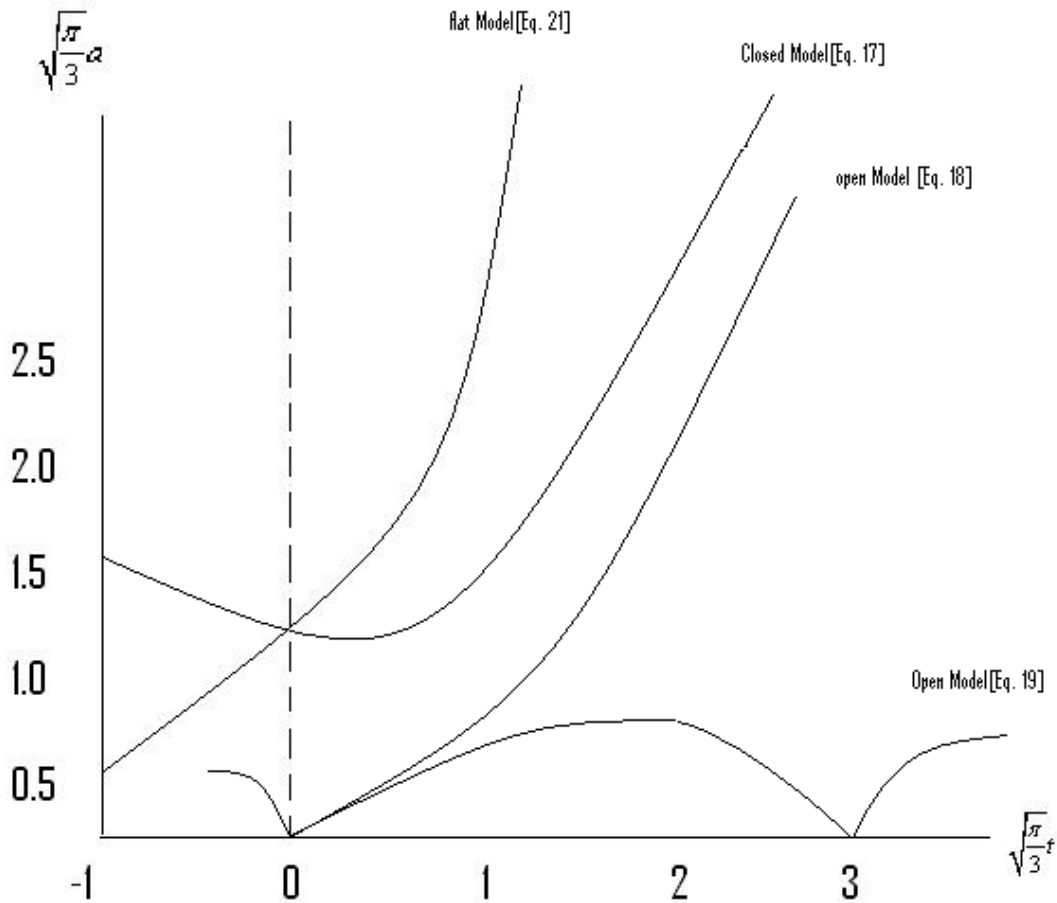

The differential equation for General Lemaitre model both with matter density and cosmological term cannot be integrated in terms of elementary functions. Thus, Let the universe begins with a bang. In the early universe the mass density is very large and we can neglect $\Lambda$; **Friedmann universe**.

As the universe expands, mass density decreases, cosmological term will become more important.

In the Friedmann cases of negative or zero curvature the expansion and the decrease in mass density are monotonic, and, hence $\Lambda$, if any, will ultimately Dominate the behavior of the universe. Thus the universe gradually becomes empty Lemaitre universe of negative or zero curvatures. In case of negative Curvature with $\Lambda < 0$, the expansion of Lemaitre universe will Stop at some later time [Fig: open model eq$^n$ 19]; the universe reverses and finally ends up in contracting Friedmann universe of negative curvature.

In case of Friedmann universe of positive curvature, the mass density reaches a minimum when one –half the period has elapsed. So, $\Lambda$ will dominate only if it is sufficiently large compared to minimum mass density. The critical value is $\Lambda = ( \ /2 \ )$.

If $\Lambda > \Lambda$ , then what began as closed Friedmann universe turn into a closed expanding Lemaitre universe [Fig: closed model eq$^n$ 17]. The transition depends upon value of $\Lambda$. For $\Lambda = \Lambda$ , the transition is never completed , the universe gets stuck at a constant value of a, i.e. a = const. $= \frac{\phantom{x}}{\Lambda}$.

The static universe with this constant value of a is known as Einstein universe. However, the equilibrium at a $= \frac{\phantom{x}}{\Lambda}$ is unstable. Any perturbation in a leads either to monotonic expansion(an expanding Lemaitre universe) or monotonic contraction (a contracting Friedmann

universe). A typical behavior of a(t) for ∧>∧ is shown below:

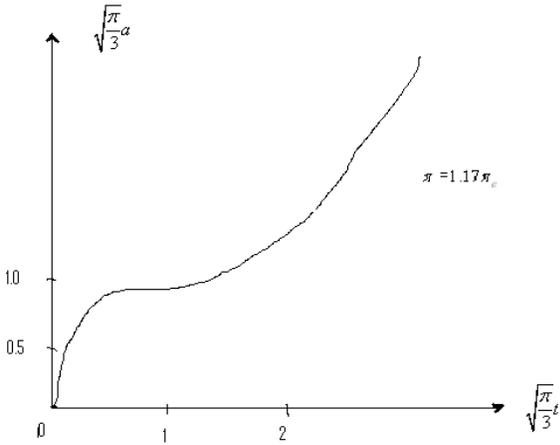

One can therefore conclude that the general Lemaitre model with a non zero mass density and ∧< 0 must necessarily be of oscillating type. If this universe was monotonic, it would gradually approach a monotonic empty model, which, as we have seen, does not exist.

So we have the basis of three observational facts:

The universe is in a state of uniform expansion.

The universe is filled with photons that come from black body background radiation.

The universe is isotropic on large scales beyond 1000 Mpc to construct or to accept, from GR, a generic cosmological modal.

Further we take the metric in the form:

$$= \quad - \quad ( )[ \ ( \quad + \quad ) + \frac{\quad}{\quad} ]$$

Where it depends on the value of k(-1,0,1) according to whether space is hyperbolic, flat or spherical and R is the characteristic size.

In-moving coordinate system matter is at rest and $T_{ij}$ is diagonal with

$$= \quad , \quad = \quad = \quad =$$

being density and p being pressure.

A fundamental aspect of GR is that the source of gravity includes explicitly a term coming from pressure : $\rho + \frac{p}{c^2}$.

Also, analogous to Gauss Theorem, Birkhoff's theorem in GR states if the matter distribution is spherical the evolution of the radius of a given shell of matter does depend only on its internal content.

In this light consider a spherical region of radius a in a homogeneous distribution of matter. Then

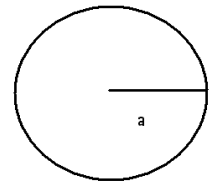

$$\frac{\ddot{a}}{a} = \frac{\ddot{R}}{R} = -\frac{4\pi G}{3}\left(\rho + \frac{3p}{c^2}\right)$$

The density term includes the effect of kinetic energy (E=mc²), so the energy conservation can be written inside the volume of sphere. Elementary thermodynamics gives

$$d(\rho c^2 V) = d(E) = -p\, dV$$

Or,
$$\dot{\rho} = -3\left(\rho + \frac{p}{c^2}\right)\frac{\dot{a}}{a}$$

Eliminating pressure between two equations we get

$$\left(\frac{\dot{a}}{a}\right)^2 = \frac{8\pi G \rho}{3} - \frac{K}{a^2}$$

The last term correspond to constant of integration associated to total energy of sphere. Its value depends on initial conditions. It expresses a link between geometry and the material content of the universe. The form of equation is independent of a, therefore it hold also for R(t)

$$\left(\frac{\dot{R}}{R}\right)^2 = \frac{8\pi G \rho}{3} - \frac{K}{R^2(t)}$$

K is replaced by k if R-N-metric.

We need a equation of state for the context of the universe to specify R(t) completely. We often find two cases where p=0 (dust) and $p = \frac{1}{3}\rho c^2$ (radiation).

Vacuum is a particular medium. Consider a piston with vacuum in it, and also assume simple vacuum is present outside. The energy inside piston is

$$E = \rho_v c^2 v$$

And if volume changes by a small account, net changes in energy are:

$$dE = d(\rho_v c^2 v) = -p_v dv$$

So, the equation of state is

$$p_v = -\rho_v c^2$$

Hence for $p \geq 0, \rho \geq 0$, there is only one solution $p=0, \rho = 0$. This is the equation of state of vacuum. This can directly be introduced in the equation governing R(t) by introducing a constant

$$\Lambda = 8\pi G \rho_v$$

Such a term is called Cosmological constant and has been historically introduced by listening by Einstein to modify his theory. So we have

$$\left(\frac{\dot{R}}{R}\right)^2 = \frac{8\pi G \rho}{3} - \frac{Rc^2}{3} + \frac{\Lambda}{3}$$

$$\dot{\rho} = -3\left(\frac{p}{c^2} + \rho\right)\frac{\dot{R}}{R}$$

### REMARKS

Obsevation favors acceleration universe indicating existence of "dark energy" which can be identified with vacuum energy ar called "quintessence".

Ratras and Peebles 1988 Phy Rev. D. 37, 3406

Peebles and Ratra 2003 rev. Mod. Phys. 75,559

Coldwell et al 1998 Phy. Rev. Let., 88, 1582

Some "reduced" quantities":

The Habble parameter $H = \dfrac{\dot{R}}{R}$

The density parameter $\Omega_M = \Omega = \dfrac{8\pi G \rho}{3H^2}$

The deceleration parameter $q = -\dfrac{R\ddot{R}}{\dot{R}^2}$

The (reduced) cosmological constant $\Omega_\lambda = \lambda = \dfrac{\Lambda}{3H^2}$

The curvature parameter $\alpha = -\Omega_k = \dfrac{kc^2}{H^2 R^2}$

In this light we have 
$$\Omega_M + \Omega_k + \Omega_\lambda = 1$$
$$\text{or} \quad \alpha = \Omega + \lambda - 1$$

So that "radius of the universe "can" be written as $R = \dfrac{C}{H}\dfrac{1}{\sqrt{|\alpha|}}$

**References:**

1. Pandey S N, Int. J. Theo. Phys., 22(1983), 209.

2. Pandey S N, Int. J. Theo. Phys., 27(1988), 695.

3. Grishchuk L P, Sov. Phys. JETP, 40(1975), 409.

4  Grishchuk L P, Sov. Phys. Usp., 20(1977), 319.

5. Pandey S N, Nuoro Cimento, 116(2001), 327.

6. Gravitation: Phoebix Pub. Co. 1995.